\documentstyle[sprocl,mydef]{article}
\begin{document}


\title{SIGNIFICANCE OF THE SPINORIAL BASIS IN THE~QUANTUM
THEORY\,$^\ast$}

\author{VALERI V. DVOEGLAZOV\,$^\dagger$}

\address{
Escuela de F\'{\i}sica, Universidad Aut\'onoma de Zacatecas \\
Antonio Doval\'{\i} Jaime\, s/n, Zacatecas 98068, ZAC., M\'exico\\
Internet address:  VALERI@CANTERA.REDUAZ.MX}


\maketitle\abstracts{Problems connected with a choice of the spinorial
basis in the $(j,0)\oplus (0,j)$ representation space are discussed.  As
shown it has profound  significance in the relativistic quantum theory.
{}From the methodological viewpoint this fact is related with the important
dynamical role played by space-time symmetries for all kind of
interactions.}


\begin{footnotesize}
\rightline{Asegura el centro, el cielo y la tierra}
\rightline{\hspace*{-12mm} encontrar\'an sus lugares adecuados.}
\rightline{Chinise saying}
\end{footnotesize}

\smallskip

My contribution deals with  mathematical constructions in
the $(j,0)\oplus (0,j)$
representation space of the extended Lorentz group
including reflections. While
the Dirac theory ($j=1/2$) found  an overall application
for describing the particle world,
the possibility of other  models in  such representations
and their possible relevance to
other kind of interactions (as well as  to electromagnetic
interaction) appear not to have been realized completely
until now.  To my knowledge, these topics were in the area of
interests of Prof. E. P. Wigner~\cite{Wigner,Wigner2} and
of a number of other eminent physicists.
The explicit constructions
of the quantum field theory of the Bargmann-Wightman-Wigner
type,~$^{1b}$ presented in refs.~\cite{DVA,DVA1}, provide
deeper insights to the essence of the problem, particularly, in the
physics of neutral particles. We start from a  general set of postulates,
which are assumed to be valid for any relativistic quantum theory:

\smallskip

$\bullet$\,For arbitrary $j$ the right $(j,0)$ and the left $(0,j)$ handed
spinors transform in the following ways (according to the Wigner's
rules~\cite{Wigner}):
\begin{eqnarray}
\phi_{_R} (p^\mu)\, &=& \,\Lambda_R (p^\mu \leftarrow
\overcirc{p}^\mu)\,\phi_{_R} (\overcirc{p}^\mu) \, = \, \exp (+\,{\vec J}
\cdot {\vec \varphi}) \,\phi_{_R} (\overcirc{p}^\mu)\quad,\\
\phi_{_L} (p^\mu)\, &=&\, \Lambda_L (p^\mu \leftarrow
\overcirc{p}^\mu)\,\phi_{_L} (\overcirc{p}^\mu) \, = \,  \exp (-\,{\vec J}
\cdot {\vec \varphi})\,\phi_{_L} (\overcirc{p}^\mu)\quad.\label{boost0}
\end{eqnarray}
$\Lambda_{R,L}$ are the matrices for Lorentz boosts; ${\vec J}$ are
the spin matrices for spin $j$; ${\vec \varphi}$ are parameters of the
given boost.  If we restrict ourselves to the case of bradyons they are
defined by formulas (3) of ref.~$^{3a}$.

$\bullet$\,\, $\phi_{_L}$ and $\phi_{_R}$ are the eigenspinors of the
helicity operator $({\vec J}\cdot {\vec n})=({\vec J}\cdot {\vec p})/\vert
{\vec p}\vert$:
\begin{equation}
({\vec J}\cdot {\vec n})\,\phi_{_{R,L}}
\,=\, h \,\phi_{_{R,L}}
\end{equation}
($h = -j, -j+1,\ldots j$  is the
helicity quantum number).

$\bullet$\, The relativistic dispersion relations
$E= \pm \sqrt{ {\vec p}^{\,2} +m^2}$  are valid for observed particle
states.

\smallskip

On the basis of these postulates and with taking
into account the Ryder-Burgard
relation~\cite{DVA}, that states that ``when a particle is at rest,
one cannot define its spin as either left- or right-handed", we derive:

$\bullet$\, The mathematical generalization of the Dirac equation
in the $(1/2,0)\oplus (0,1/2)$
representation space (momentum representation):
\begin{equation}
\left [a \,{\hat p \over m} + b \,{\cal T} \,S^c_{[1/2]}
- {\cal T}\right ]\Psi (p^\mu) = 0\quad,\quad a^2 +b^2 =1\quad;
\end{equation}
$S^c_{[1/2]}$ is the charge conjugation operator;
${\cal T} \equiv  \cos \alpha -i\gamma^5 \sin \alpha$; $\alpha$
is the phase factor
entering in the generalized Ryder-Burgard relation:\,\cite{DVO1}
\begin{equation}
\phi_{_R} (\overcirc{p}^\mu) \,=\,  a \,e^{i\alpha_{\pm}} \,\phi_{_L}
(\overcirc{p}^\mu) + i\,b \,e^{i\beta_{\mp}}\, \Theta_{[1/2]}
\,\phi_{_L}^* (\overcirc{p}^\mu)\quad.
\end{equation}

$\bullet$\,The particular model with $a = 1$, $b=0$ and $\alpha = \pm {\pi
\over 2}$.  The remarkable feature of this model is  the necessity of
introducing two field functions.  In the opposite case, when one has only
one field function having plane-wave expansion, we cannot construct the
Lagrangian in the coordinate representation.  The physical system is
described  by a set of two equations:
\begin{equation}
\left [\gamma^5 \gamma^\mu \partial_\mu +m \right ]\Psi^{(1)} (x) = 0\quad,
\quad\left [\gamma^5 \gamma^\mu \partial_\mu -m \right ]\Psi^{(2)} (x) =
0\quad, \label{eq2}
\end{equation}
connected with the Dirac equation by
unitary transformations.  The physical consequences of  the model are the
following: 1)\,One can obtain  charged particles, in
fact,  Dirac fermions, but  neutral particles too. 2)\,One can describe
bosons in the framework of the $(1/2,0)\oplus (0,1/2)$ representation
space. 3)\,There is a  puzzled physical ``excitation" with $E\equiv 0$,
$Q\equiv 0$ and $(W\cdot n) \equiv 0$. 4)\,The Feynman-Dyson propagators
for fields $\Psi^{(1)}$ and $\Psi^{(2)}$ are not equal to the Wick
propagators ({\it cf.} with a $j=1$ case). 5)\,The question of
self-energy contributions and, hence,  mass splitting between proper-mass
states $\vert +m >$ and $\vert -m >$ induces development of  adequate
methods for higher-order calculations ({\it e.g.} in the spirit
of~\cite{Kik}).

$\bullet$\,The connection with the Majorana-Ahluwalia Construct,
which is based on the use of the 4-spinors of type-II:
\begin{equation}
\lambda(p^\mu)\,\equiv \pmatrix{
\left ( \zeta_\lambda\,\Theta_{[j]}\right )\,\phi^\ast_{_L}(p^\mu)\cr
\phi_{_L}(p^\mu)} \,\,,\quad \rho(p^\mu)\,\equiv \pmatrix{
\phi_{_R}(p^\mu)\cr
\left ( \zeta_\rho\,\Theta_{[j]}\right )^\ast
\,\phi^\ast_{_R}(p^\mu)} \,\,\quad .\label{sp-dva}
\end{equation}
They can be introduced for any $j$. They are not in the helicity
eigenstates\,\footnote{Let us still not forget that the helicity quantum
number is not  a ``good" quantum number for massive particles.} due to the
properties of the Wigner time-reversal  operator:  $\left
(\Theta_{[j]}\right )_{\sigma,\,\sigma^\prime} = (-1)^{j+\sigma}
\delta_{\sigma^\prime ,\, -\sigma}$\, , $\Theta_{[j]} {\vec J}
\Theta^{-1}_{[j]} = -{\vec J}^*$\, . Phase
factors $\xi_\lambda$ and $\xi_\rho$ are fixed by the conditions of
self/anti-self $\theta$-conjugacy. On the base of  consideration of the
4-spinors of the second type one can still  deduce both the equations
(\ref{eq2})\,\footnote{Spinors
satisfying equations (\ref{eq2}) in the momentum representation
are in helicity eigenstates. They are connected with $\lambda^{S,A}$
as follows:~\cite{DVO2}
\begin{eqnarray}
\Upsilon_+ (p^\mu) = \left [\pm\frac{1+\gamma_5}{2} \lambda^{S,A}_\downarrow
+ \frac{1-\gamma_5}{2} \lambda^{S,A}_\uparrow\right ]\, , \,
\Upsilon_- (p^\mu) = \left [\mp\frac{1+\gamma_5}{2} \lambda^{S,A}_\uparrow
+ \frac{1-\gamma_5}{2} \lambda^{S,A}_\downarrow\right ],\\
{\cal B}_+ (p^\mu) = \left [\mp\frac{1+\gamma_5}{2} \lambda^{S,A}_\downarrow
+ \frac{1-\gamma_5}{2} \lambda^{S,A}_\uparrow\right ]\, , \,
{\cal B}_- (p^\mu) = \left [\pm\frac{1+\gamma_5}{2} \lambda^{S,A}_\uparrow
+ \frac{1-\gamma_5}{2} \lambda^{S,A}_\downarrow\right ].
\end{eqnarray}
Helicity eigenspinors are connected with $\rho^{S,A}$ spinors
in the similar fashion.
The arrows $\uparrow\downarrow$ should be referred to `the chiral
helicity'  introduced in ref.~\cite{DVA1}.}
and the ``mad" forms of the
Dirac equation:\,\footnote{We assumed that  ``parity violation"
is not explicit in the meaning
of ref.~\cite{Barut}. See formulas (9,48a,48b) in ref.~\cite{DVA1}.}
\begin{eqnarray}
i\gamma^\mu \partial_\mu \lambda^S (x) - m \rho^A (x) &=& 0\quad,\quad
i\gamma^\mu \partial_\mu \rho^A (x) - m \lambda^S (x) =
0\quad,\label{eq3}\\
i\gamma^\mu \partial_\mu \lambda^A (x) + m \rho^S (x)
&=& 0\quad,\quad i\gamma^\mu \partial_\mu \rho^S (x) + m \lambda^A (x) =
0\quad.\label{eq4} \end{eqnarray}
The physical states described by Eqs.
(\ref{eq3},\ref{eq4}) can possess axial charge.  $\lambda$ and $\rho$
spinors are closely related with the asymptotically chiral massive fields
introduced by Ziino and Barut.~\cite{Barut} Neutrino and its antineutrino
can be considered to coincide in the framework of this theory. The
question is what dynamical behaviour do we have,  the Dirac-like or the
Majorana-like. The answer depends on relations between type-II spinors
and their parity-conjugates. Finally, let us note that at-rest spinors
$\Upsilon_\pm$, \, ${\cal B}_\pm$ are connected by unitary transformations
with the Ahluwalia's type-II spinors:  \,$\Upsilon_{\pm 1/2}
(\overcirc{p}^\mu) = U \lambda^S_{\uparrow\downarrow} (\overcirc{p}^\mu) =
-\gamma_5 U \lambda^A_{\uparrow\downarrow} (\overcirc{p}^\mu)$\,\, ${\cal
B}_{\pm 1/2} (\overcirc{p}^\mu) = U \lambda^A_{\uparrow\downarrow}
(\overcirc{p}^\mu) = -\gamma_5 U \lambda^S_{\uparrow\downarrow}
(\overcirc{p}^\mu)$\, .
The transformation matrix is
\begin{eqnarray}
U = \pmatrix{\Xi^{-1}_{[1/2]} \Theta^{-1}_{[1/2]} & 0\cr
0 & 1\cr}\quad.
\end{eqnarray}

$\bullet$\, Similarities and differences in results  for the $(1,0)\oplus
(0,1)$ (and higher-spin representations)  space. {\it E.g.},
self/anti-self conjugate spinors do not exist for spin-1 in the
considered model. The equations for $\lambda^{S,A}
(p^\mu)$ and $\rho^{S,A} (p^\mu)$  ``6-spinors" are connected with the
Weinberg equation of 1964.

Finally, I would like to draw your attention to the problem of the claimed
longitudity of
the antisymmetric tensor field (transformed also according
to the $(1,0)\oplus (0,1)$
representation) after quantization.
In my opinion, such a situation is quite unacceptable since it
produces speculations on the violation of the Correspondence Principle and
on the contradiction with the Weinberg theorem $B-A=h$.   Assumingly, this
question has relations with the problem at hand. The  answer which we
achieved is:  ``the  queer reduction of degrees of freedom" is happening
because of  application of the generalized Lorentz condition.  What is
this good for? I mean, is it necessary to make a non-linear realization of
a scalar field from an antisymmetric tensor field?   As a result of
analysis of the theory of the antisymmetric tensor field with transversal
components, I support Sachs and Barut ideas of the unified treatment  of
field and sources in the gauge theories.

Concluding, we note that the $(j,0)\oplus (0,j)$ representation space has
a rich structure which  should be analyzed carefully  with the aims of
probable applications for the unification of all types of interactions.
Details of the formalism presented here could be found
in refs.~\cite{DVA1,DVO1,DVO2}.

\section*{Acknowledgments}

I acknowledge the help of Prof. D. V. Ahluwalia, whose articles and
preprints are the basis for this work. The valuable advice of Prof. A.
F. Pashkov was helpful in realizing ideas presented here and in my
previous publications. I thank  organizers of the Wigner Symposium for the
possibility of presentation of ideas which are, in my opinion, the logical
continuation of  pioneer researches of E. P. Wigner.   I dedicate
this paper to the memory of  Prof. A. O. Barut, Prof. M. A. Markov and
Prof.  E. P. Wigner.


\footnotesize{
\,$^\ast$\, Reported at the III Reuni\'on Anual de la DFMG
(3-4/05/95), the IX Reuni\'on Anual de la DPC  de la Sociedad Mexicana de
F\'{\i}sica (21-23/06/95), M\'exico city and at the IV Wigner
Symposium (7-11/08/95), Guadalajara, M\'exico.\quad \,$^\dagger$\, On
leave of absence from {\it Dept. Theor.  \& Nucl. Phys., Saratov State
University, Astrakhanskaya ul., 83, Saratov\, RUSSIA.} Internet address:
dvoeglazov@main1.jinr.dubna.su }

\normalsize{

}

\end{document}